\documentclass[copyright]{eptcs}
\providecommand{\event}{MBT 2014} % Name of the event you are submitting to
\usepackage{graphicx}
\usepackage{wrapfig}
\usepackage{caption}
\usepackage{subcaption}
\usepackage[nottoc,numbib]{tocbibind}

\title{Using Formal Specifications to Support Model Based Testing\\ASDSpec: A Tool Combining the Best of Two Techniques}
\author{A.P. van der Meer \institute{Nspyre} \email{arjan.van.der.meer@nspyre.nl} \and R. Kherrazi \institute{Nspyre} \email{rachid.kherrazi@nspyre.nl} \and M. Hamilton \institute{Nspyre} \email{marc.hamilton@nspyre.nl}}
\def\titlerunning{Using Formal Specifications to Support Model Based Testing}
\def\authorrunning{A.P. van der Meer \& R. Kherrazi \& M. Hamilton}
\newcommand{\csharp}{C$^\sharp$ }
\newcommand{\ASDSpec}{ASDSpec }
\begin{document}

\maketitle

\begin{abstract}
Formal methods and testing are two important approaches that assist in the development of high quality software. For long time these approaches have been seen as competitors and there was very little interaction between the two communities. In recent years a new consensus has developed in which they are seen as more complementary. In this report we present an approach based on the ASD(Analytical Software Design) suite by Verum® and the Microsoft® Spec Explorer Model Based Testing(MBT) tool. ASD is a model-based design approach that can produce verified software components that can be combined into complete systems. However, ASD cannot verify existing components, nor complex component interactions involving data transfers. We have developed a tool that allows us to convert ASD models to Spec Explorer, allowing us to do more complete verification of software systems using dynamic testing at little additional cost and effort. We demonstrate this by applying our approach to an industrial-size case study.
\end{abstract}


\section{Introduction}
\label{sec:Introduction}
As with any kind of construction, reliability is of utmost importance in software engineering. To achieve this, Verum®~\cite{VerumSite} has created Analytical Software Design (ASD), a component based  design method. By using this method, software can be constructed that is guaranteed to implement specified interfaces and protocols and is deadlock-free. It is based on the decomposition of a system in different components modeled by state machines. These models can then be used to generate formally verified code that can be integrated to construct complete and correct industrial scale systems. However, in practice, many software systems require the use of third party and/or legacy components that have not been verified in addition to generated components. The ASD method cannot guarantee correctness for these external components nor for the integrated system, because it only performs static verification on a per-component bases and no dynamic testing nor testing of composed systems. To remedy this, we propose to use Model-Based Testing (MBT) to verify if the external components implement the behavior of the interfaces defined in the relevant ASD models, and if the complete system, including the external components, fulfills all requirements.
We implemented a prototype of this approach that transforms ASD interface models to Microsoft® Spec Explorer\cite{SpecExplorerSite} models. Spec Explorer allows us to generate automated test cases that explore the expected interface behavior of any component, and enables us to find violations of the requirements with a minimum of manual effort. Furthermore, in addition to the verification ASD provides, Spec Explorer can use data testing to validate that the system implements desired behavior by linking inputs to observable events. Finally, Spec Explorer allows us to use model composition to test complete system, while also supporting data abstraction and combination~\cite{DBLP:conf/issre/VishalKKM12} to reduce the state space to acceptable levels. In contrast, the facilities in ASD for handling data are very limited. By reusing ASD interface models in Spec Explorer, this novel method gives us the full usage of all benefits of this powerful MBT tool (e.g. data abstraction to reduce state space and overcome state explosion and models decomposition) without facing issues related to creation of complex test models.
In this paper, we will first introduce ASD and Spec Explorer in more detail in Section~\ref{sec:Preliminaries}. We will then discuss the fundamental similarities that we use in our tool in Section~\ref{sec:ImplementationConcept}. In Section~\ref{sec:ImplementationAppproach} we describe some technical details of the implementation. In Section~\ref{sec:CaseStudy}, we describe a study into the effectiveness of our tool and present some empirical results. Section~\ref{sec:RelatedWork} we discuss some related work, and finally conclusions in Section~\ref{sec:Conclusions}.
\section{Preliminaries}
\label{sec:Preliminaries}
\subsection{Analytical Software Design (ASD)}
\label{sec:ASD}
\begin{figure}[hbp]
  \begin{subfigure}[t]{0.3\textwidth}
  \includegraphics[width=\textwidth]{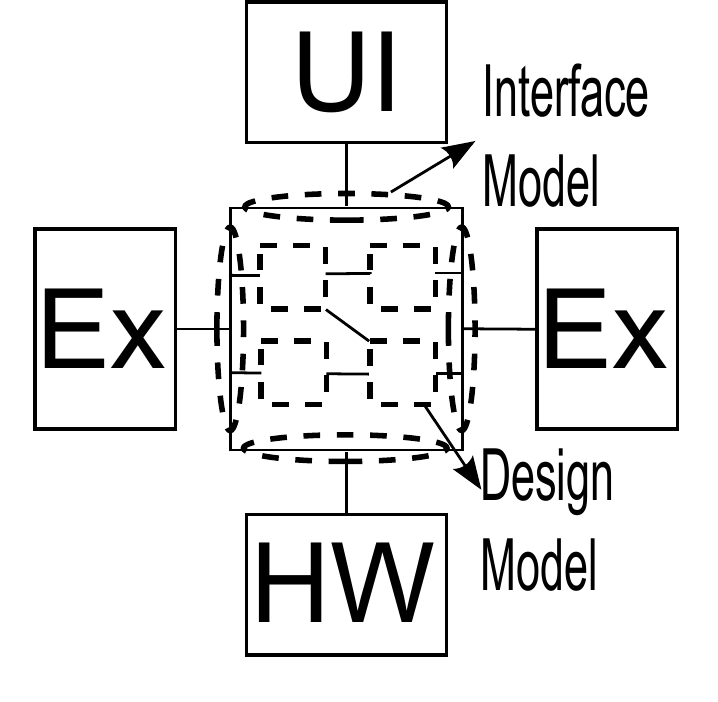}
  \caption{ASD specifications consist of design and interface models}  
  \label{fig:ASDModels}
  \end{subfigure}
  \begin{subfigure}[t]{0.3\textwidth}
  \includegraphics[width=\textwidth]{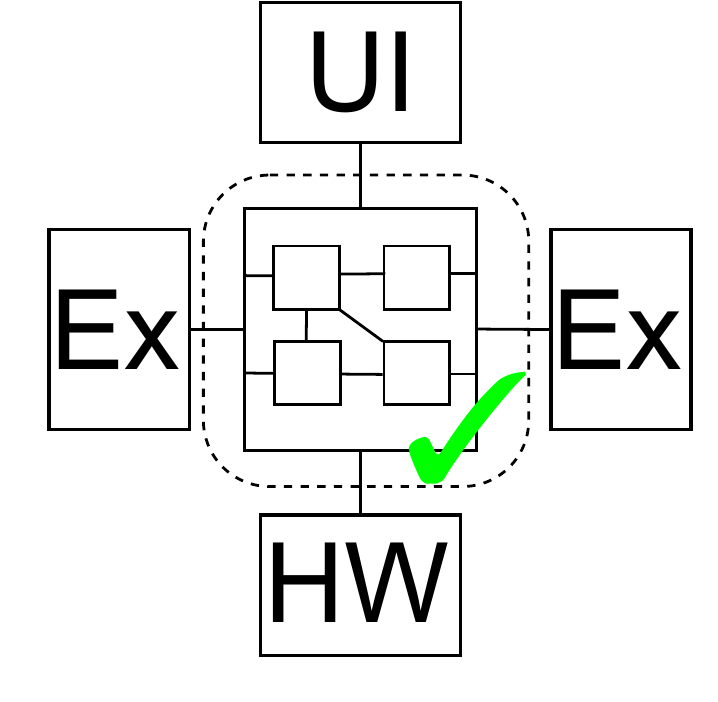}
  \caption{ASD guarantees correctness of generated code using static verification}
  \label{fig:ASDsystem}
  \end{subfigure}
  \begin{subfigure}[t]{0.3\textwidth}
  \begin{center}
  \includegraphics[width=\textwidth]{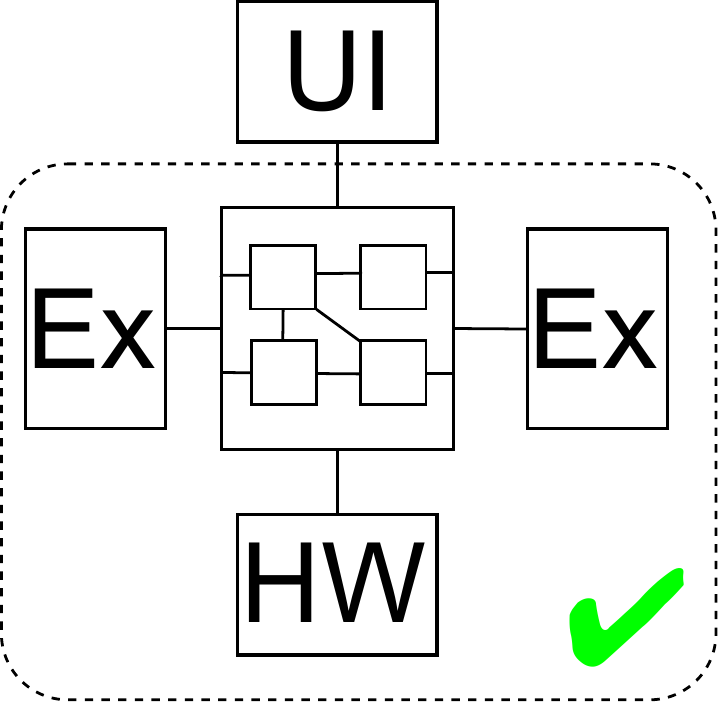}
  \caption{Spec Explorer verifies complete system using dynamic testing}
  \label{fig:SpecExplorersystem}
  \end{center}
  \end{subfigure}
\end{figure}
The ASD technology is developed by Verum with the primary aim of supporting the development of complex embedded software and increasing its quality. In the ASD approach, software systems are envisioned as a collection of components that communicate via interfaces. Figure~\ref{fig:ASDModels} shows the overall structure of an ASD specification. To develop a new piece of software using ASD, we first have to specify one or more desired interfaces as interface models. As Figure~\ref{fig:ASDModels} illustrates, these interfaces can involve hardware drivers (HW), user interfaces (UI) and other external components (EX). In addition to defining the operations, referred to as events, that the interface supports and the responses it can give, an interface model contains a state machine that defines the protocol that has to be used to access the interface. One of the main features of the ASD tool set is that these protocols are strictly enforced, so an ASD-generated software component can never violate them. once the interface models have been defined, design models can be created that describe how to implement the interfaces. This is also done using state machines. An example of an ASD state machine is shown in Figure~\ref{fig:ASDScreenshot}, with a visualization of the state machine on the left and a fragment of the definition created by the designer on the right. As shown, a state machine is constructed using a table layout, where for each state and for each possible event is defined how the component should respond. If the event is expected, the response can consist of changing the state of the component and sending responses to the calling component. Otherwise, the event is declared illegal, and if it occurs in the given state, this results in a failed verification. Overall, as shown in Figure~\ref{fig:ASDsystem}, this means ASD is focused on the core part of the system, while external components or hardware are not verified.

\noindent
Once the interface models are complete, the next step is to create a design model that combines all interfaces. Like an interface model, a design model contains a state machine, but in this case it implements the behavior of the desired software component. To do this, a design model can refer to other interfaces, using them to provide some required functionality. When the design model is complete, ASD will verify that any used interfaces are always invoked according to their specifications. If the implementations of the interfaces are also generated with ASD, the resulting system is guaranteed to be correct with respect. However, if the implementation is actually third-party or legacy code, the component can be incorrect and invoke invalid events or respond incorrectly. In such cases, the generated system does not know how to respond and stops functioning. Verum suggests remedying this by introducing a so-called Armour layer between ASD systems and external components that filters any undesired communication, but this still means correct functioning of the system cannot be guaranteed. This is a fundamental limitation of the static testing concept used by ASD: anything not described directly by a model cannot be verified. In the case of larger systems with many components, ASD is also limited in its ability to verify complex interactions, because the number of states grows beyond the static testing it can do. 
\subsection{Spec Explorer}
\label{sec:SpecExplorer}
Spec Explorer is an extension to Microsoft Visual Studio intended to provide support for MBT. To use Spec Explorer, we first have to define a model that describes the expected behavior of the system to be tested. This model consists of one or more \csharp classes enhanced with modeling annotations. The model is used to compactly define the possible behavior of the system. Once the model is complete, we can apply testing strategies to generate test cases, which can be executed directly to see if the implemented system meets our expectations. In each model class, methods can describe behavior the system under test should implement. As any ordinary \csharp method, they can update variables, invoke other methods and return values. By computing the effects of each method, Spec Explorer can construct a state space containing all required behaviors. Each path in this state space represents a possible test, a sequence of steps that the system under test should be able to follow. Spec Explorer offers a range of strategies to select a representative sample of paths, based on for example data coverage. As shown in Figure~\ref{fig:SpecExplorersystem}, because events are invoked on the system as a whole, this means not only the generated code but also external and even hardware components are involved in the tests.

\begin{figure}[hbp]
\fbox{\includegraphics[width=0.25\textwidth]{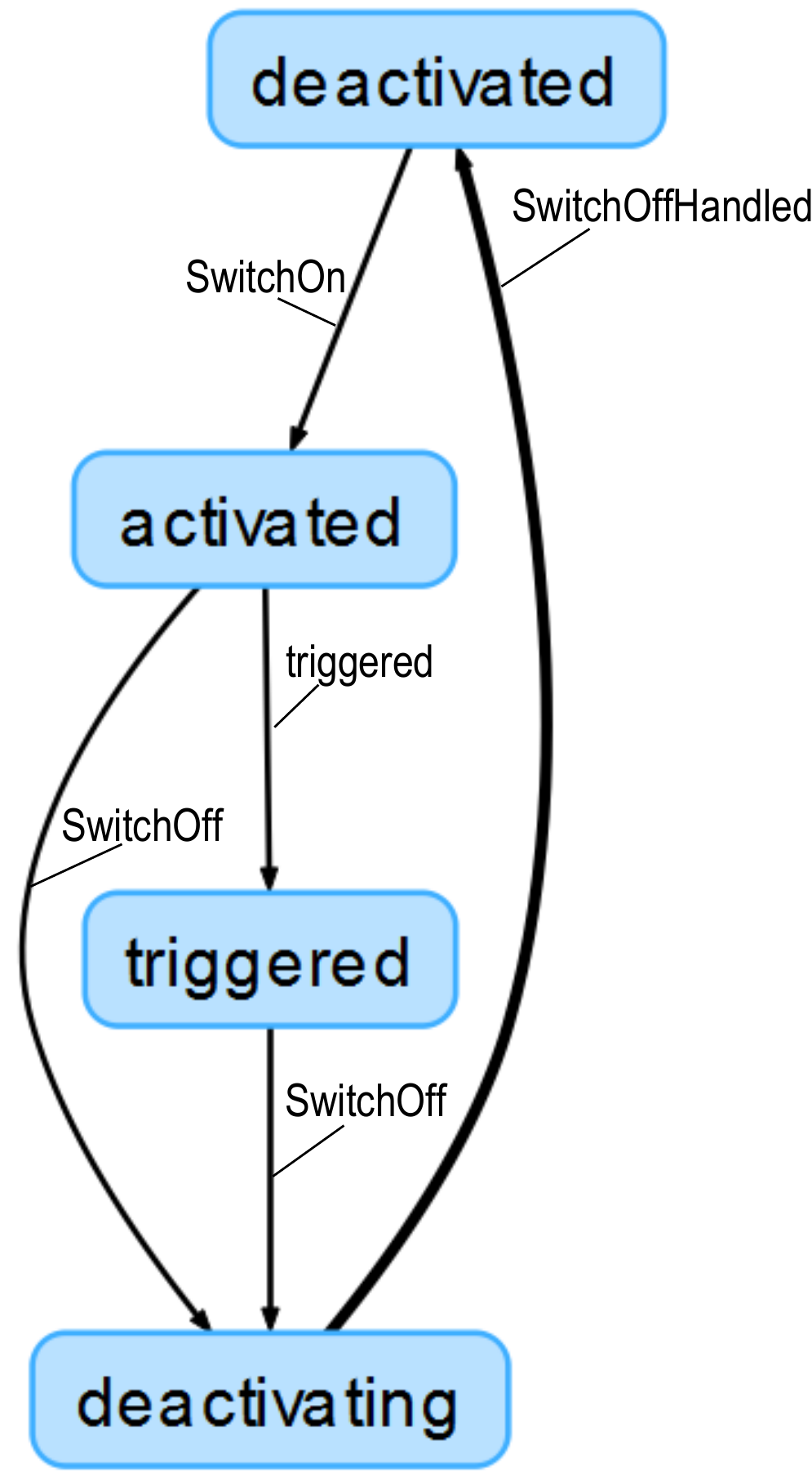}}
\includegraphics[width=0.75\textwidth]{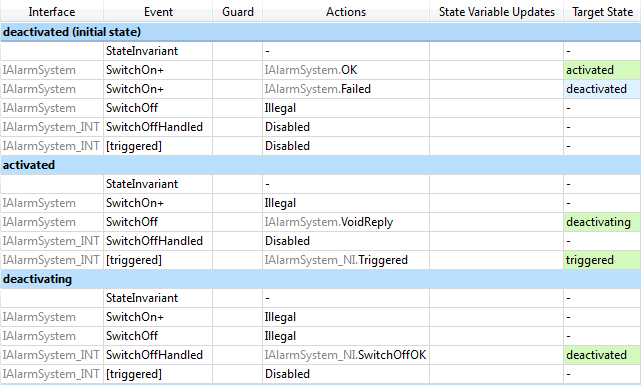}
\caption{Fragment of an ASD state machine definition and corresponding state diagram}
\label{fig:ASDScreenshot}
\end{figure}

\begin{figure}[p]
\fbox{\includegraphics[width=0.35\textwidth]{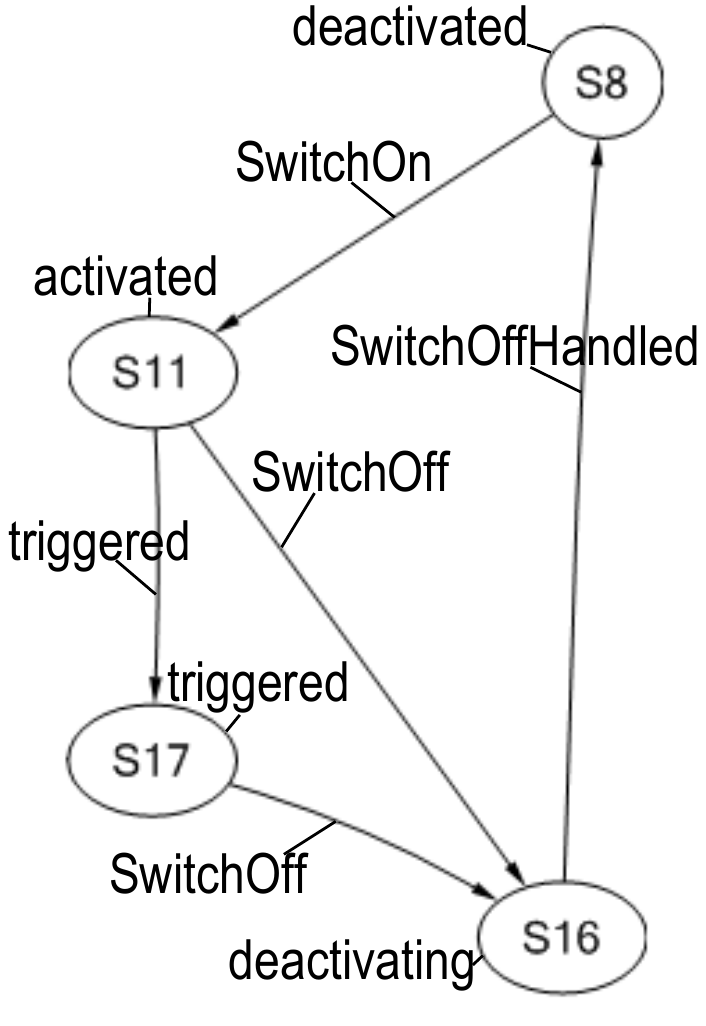}}
\includegraphics[width=0.65\textwidth]{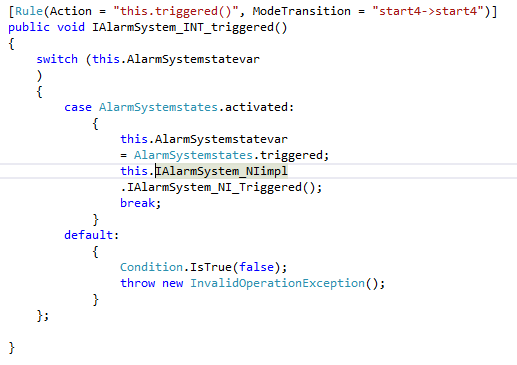}
\caption{Fragment of a generated Spec Explorer model and part of the corresponding state diagram}
\label{fig:SpecExplorerScreenshot}
\end{figure}

\begin{figure}[hbp]
\fbox{\includegraphics[width=0.35\textwidth]{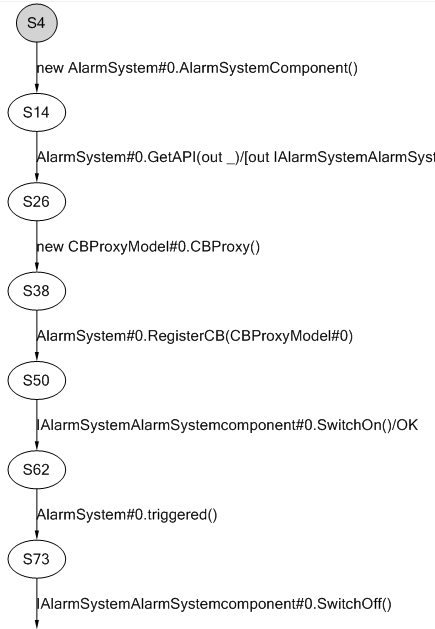}}
\includegraphics[width=0.65\textwidth]{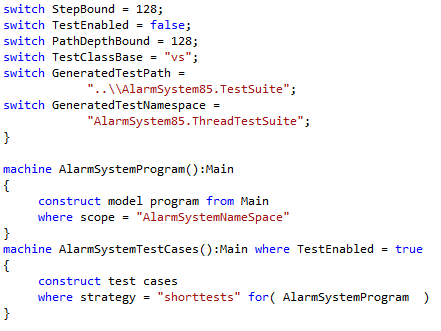}
\caption{Fragment of a CordScript script and part of generated test suite}
\label{fig:CordScriptScreenshot}
\end{figure}

How the system under test should execute the steps is described by the model annotations. Using the “TypeBinding” annotation, we relate each model class to an implementation class. Individual methods are linked to their counterparts using a “Rule” annotation. An example of a model is shown in Figure~\ref{fig:SpecExplorerScreenshot}, which shows one method of a model class together with part of the state machine defined in the model. For readability purposes, we have removed states relating only to initialization and finalization details, and added labels indicating the correspondence between Spec Explorer states and ASD states as shown in Figure~\ref{fig:ASDScreenshot}. The system under test is in this case actually an implementation of the alarm system also described in Section~\ref{sec:ASD}. This method describes how the alarm should react when it receives a \texttt{triggered} event. From the declaration of the method, we can see that it returns no value and has no parameters. In the body, we can see a switch statement that selects appropriate behavior based on the current state of the model, stored in the \texttt{AlarmSystemstatevar} variable. If the alarm is in the activated state, the first case of the switch will be used. This case specifies that the state of the model should be updated to \texttt{triggered}, and that the \texttt{IAlarmSystem\_NI\_Triggered} method of the \texttt{IAlarmSystem\_NIimpl} variable should be invoked. In all other cases, the method cannot be used. This is indicated by the \texttt{Condition.isTrue(false)} construct, which indicates to Spec Explorer that the method call is not valid and should not be used in tests. If the method would be called in the implementation during testing, this would result in an error and a failed test. In the state machine visualization, this method corresponds to the edge labeled \texttt{triggered}.
In addition to the model, Spec Explorer uses a script file in the CordScript language to define what tests should be executed. Figure~\ref{fig:CordScriptScreenshot} shows part of a CordScript file corresponding to the model shown in Figure~\ref{fig:SpecExplorerScreenshot}, implementing a basic testing strategy. A CordScript file consists of two main kinds of elements: configurations and machines. Configurations, shown in the top part of Figure~\ref{fig:CordScriptScreenshot}, define the basic parameters of the test, including the actions we are interested in and global properties like what test engine to use and how long tests can be. Machines, shown in the bottom part of Figure~\ref{fig:CordScriptScreenshot}, are used to select which tests we want to execute. In most cases, we will want tests that cover all behaviors of the system, either as one large test or a number of smaller tests. If, for example, we want to test a specific action in the system, we can use a machine to select only those tests containing that particular action. In CordScript, we can also define model compositions to test complex systems and data abstraction and combination to reduce state spaces and test sizes. In the figure, we show part of a basic configuration, consisting of a number of switches that control the test generation process and two machines, one that defines the state space of the model, \texttt{AlarmSystemProgram}, and one that defines the test strategy that we want to apply, \texttt{AlarmSystemTestCases}. The end result is a number of state machines similar to the one shown on the left in Figure~\ref{fig:CordScriptScreenshot}. The state machine represents the test case as a series of abstract steps, which can easily be translated to concrete steps which can be used by the chosen testing framework.
\subsection{Conclusions}
\begin{figure}[hbp]
\begin{center}
\includegraphics[width=0.5\textwidth]{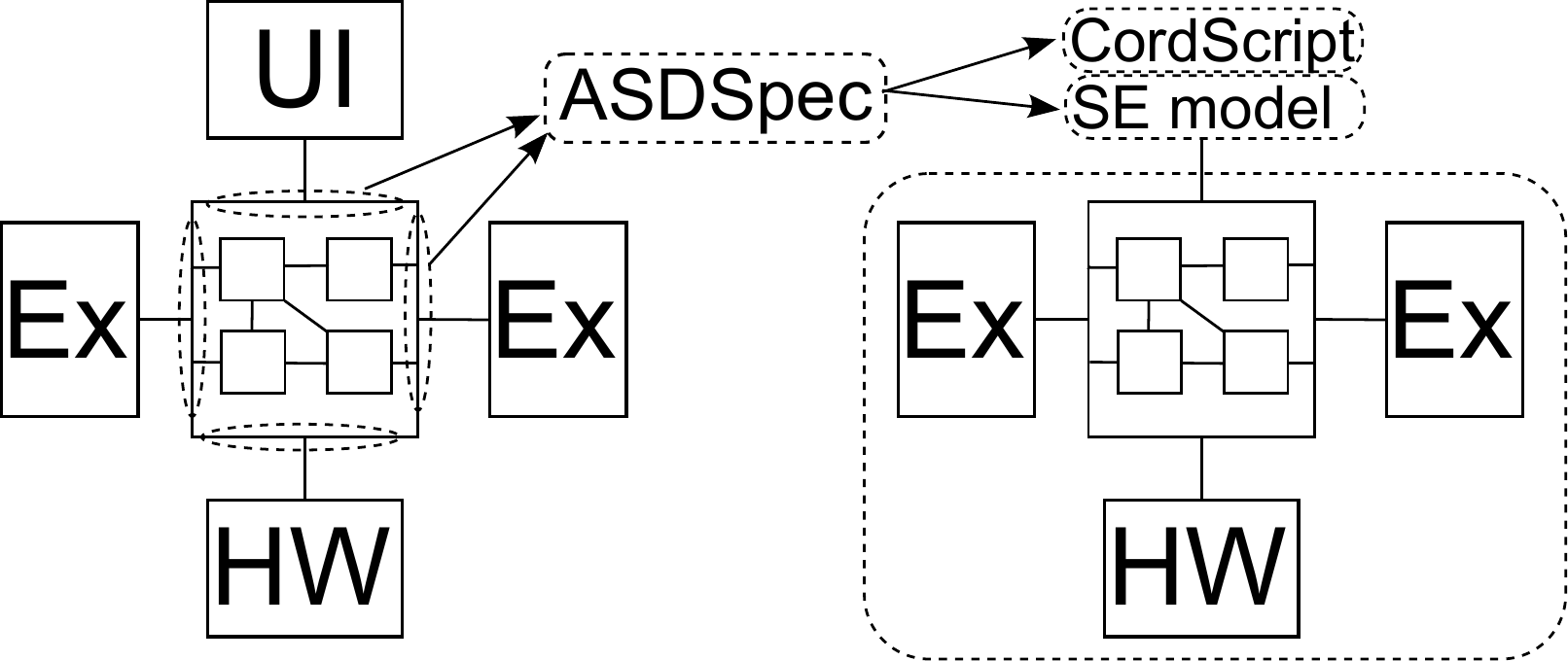}
\caption{ASDSpec converts ASD Interface Models to Spec Explorer Models and Cord Scripts}
\label{fig:SystemCombined}
\end{center}
\end{figure}

In this section, we have described ASD and Spec Explorer. By explaining the strengths and weaknesses of both approaches, we want to establish what advantages can be gained by combining the two. In Figure~\ref{fig:SystemCombined}, we show the fundamental idea in schematic form: we have a system that has to communicate with hardware(HW), other external components(EX) and a user interface(UI). During ASD verification, only the implementation of the new system is verified, while correctness of the other comments cannot be addressed. By using ASDSpec to generate Spec Explorer model based testing models, we can extend verification coverage from only the central components to the entire system. Additionally, we want to use the facilities Spec Explorer offers in composition and abstraction to test larger, more complex systems. In Section~\ref{sec:ImplementationConcept}, we will discuss the conceptual relations that make this possible in greater detail. 
\section{Implementation Concept}
\label{sec:ImplementationConcept}
In order to generate tests for a system using Spec Explorer, we need a model that describes the desired behavior and a script file that defines the kind of tests to be done. While creating these typically requires considerably little effort compared to that required to build the actual system, it is still a non-trivial process that can introduce its own errors. In the case we are testing ASD generated software, or software that has to be used in concert with ASD generated software, we already have a model that describes the required behavior. This suggests that if we can reuse this model, we can make testing with Spec Explorer easier and cheaper.

Because we cannot use ASD interface models in Spec explorer directly, this means we have to create a Spec Explorer model based on a given ASD model in a way that does not affect the semantics. In order to establish whether this is even feasible, we first have to discover what concepts ASD and Spec Explorer models share, and how they can be related. From a global perspective, we observe that both ASD and Spec Explorer are fundamentally state machine-based technologies. This means a system at all times has a well-defined state, which can change in response to triggers from the outside world. In particular, this implies that we can consider a Spec Explorer model compliant to an ASD model if the state machines involved are sufficiently equivalent in behavior. In particular, we want to make sure that the tests that are constructed by Spec Explorer can potentially contain all possible sequences of event triggers that are legal in the ASD state machine, to ensure we can reach any level of coverage we desire. On the other hand, illegal events should never occur in tests, because the behavior of the component is undefined in such cases, which means that the test can never be failed or passed.

Looking more closely, we see that in ASD, the state machines are implemented directly in the table-based language described earlier. Unfortunately, ASD is based on proprietary technology, and we do not have access to any formal definitions of the table language semantics. However, as mentioned before, the ASD tooling provides code generation support that generates executable code based on the state machines in several different languages. Based on the claims made by Verum, we can assume that the generated code for all languages are accurate representations of the intended state machine semantics. In this case, we look at the generated \csharp code, because that is the language used by Spec Explorer. We see that all ASD interface are represented by \csharp interfaces, and each ASD event is implemented as a method, declared in the appropriate interface and implemented in a \csharp class. This means that events are triggered by calling the appropriate method, which is then executed when and how the \csharp semantics dictate.

In Spec Explorer, the \csharp code of the model is used to define the potential behavior of a system. In particular, a Spec Explorer model contains one or more classes with methods, and each method describes an event that the system can respond to. By analyzing the effects of the methods on the state of the system, Spec Explorer can identify states and the transitions between them. Thus, to reconstruct a known state machine in a Spec Explorer model, we have to create one or more classes that together implement all possible events as methods, in such a way that the resulting state space matches the one in the ASD model. The former simply requires we create a method for every event. We achieve the latter by constructing the model based on an explicit state machine pattern, thus ensuring that all states are explicitly present in the model.

\begin{figure}[!htp]
\begin{center}
\includegraphics[width=0.5\textwidth]{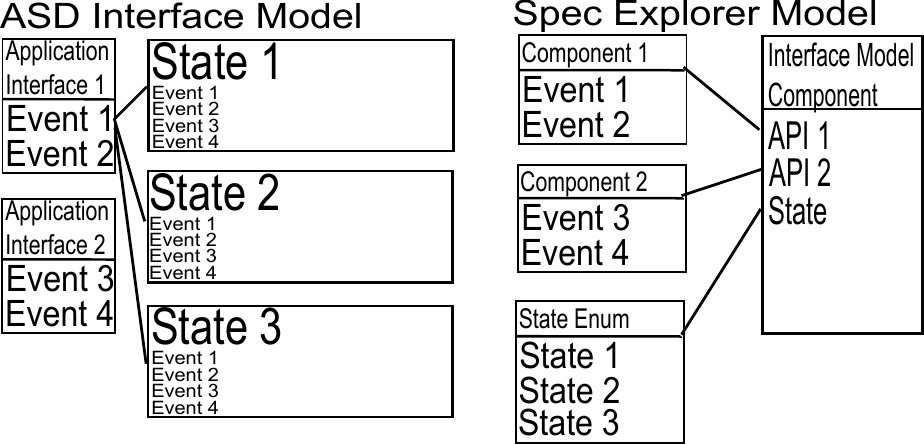}
\caption{Transformation Concept}
\label{fig:ConceptDrawing}
\end{center}
\end{figure}

In order to execute actual test runs based on generated test cases, Spec Explorer requires direct links between model events and the corresponding methods in the system under test. Because this relation is similar to the connection between ASD model events and the methods that implement them, we can use our knowledge of the ASD code generation conventions to create these links. To make this more concrete, in Figure~\ref{fig:ConceptDrawing}, we show a visualization of the structure of an ASD interface model on the left, and a visualization of corresponding \csharp code on the right. In the ASD model, we can see that the events supported by the interface model are defined in separate Application Interfaces. Separately, we can define a number of States, that for each event must describe how the system should respond to it. If the event is valid in the current state, this response consists of a transfer to a new state and possibly some actions. If the event is invalid, it is declared illegal, and the system is assumed to stop if it occurs in this state. On the \csharp side, we see that the interface model is implemented by several \csharp classes. The main class of the system is the Interface Model Component class. This class can be used to initialize the system, keeps track of the current state and provides access to implemented interfaces. Like in the ASD model, each interface contains a number of events, implement as methods. In order to trigger an event, we simply call the corresponding method. The implementation of the event will then use the current state recorded in the Interface Model Component to determine the correct behavior.
\section{Implementation Approach}
\label{sec:ImplementationAppproach}
In Section~\ref{sec:ImplementationConcept}, we have established that in order to create a Spec Explorer model based on an ASD model, we have to create a number of \csharp classes and methods that implement the corresponding state machine. To do this automatically, we choose to use a model transformation in QVTO~\cite{QVTOSite}, an Eclipse Modeling Framework(EMF) implementation of QVT Operational~\cite{QVT-specification}. Because QVTO is based on EMF~\cite{EMFSite,Steinberg:2009:EEM:1197540}, this means we have to translate ASD models into EMF form first. For this, we used an existing tool based on the XML schema provided by Verum, describing the structure of ASD models. This tool was developed by Nspyre as part of an earlier project. In the same project, a transformation was developed that abstracts from format-specific features of ASD models to a generic, more abstract representation. We use this representation as a basis for further processing.
The next step is to generate the \csharp model file and the CordScript script file. This is implemented as two QVTO transformations. The first one generates an EMF \csharp model based on a \csharp meta-model created by the MoDisco~\cite{MoDiscoSite,Bruneliere:2010:MGE:1858996.1859032} project. The second generates an EMF CordScript model based on a CordScript meta-model of our own design. Because these are both in EMF format, we then have to use templates, in our case based on the Acceleo~\cite{AcceleoSite} template engine, to create the textual representations that can be used by Spec Explorer. These templates are generic, in the sense that they can be used for all EMF \csharp and CordScript models that use our metamodels.

\section{Case Study}
\label{sec:CaseStudy}
In order to demonstrate the effectiveness of our approach, we first created a prototype, applied it to a simple model, and tested its effectiveness using bug seeding. In a next step, we looked at a larger, more complex case study. The case we choose was developed earlier by Nspyre and is based on a container terminal that could be both simulated and implemented as a demonstration model. Because this system uses external components with legacy code, it serves as a good example of the advantages of model-based testing as complement to ASD verification. In the initial project, the required Spec Explorer test models where constructed by hand, at considerable effort, as is indicated in Table~\ref{tbl:CaseResults}. This table is discussed in greater detail in Section~\ref{sec:Results}. Using \ASDSpec, we aim to reduce this cost. 

\subsection{Case description}
\begin{figure}[hbp]
\begin{center}
\includegraphics[width=0.5\textwidth]{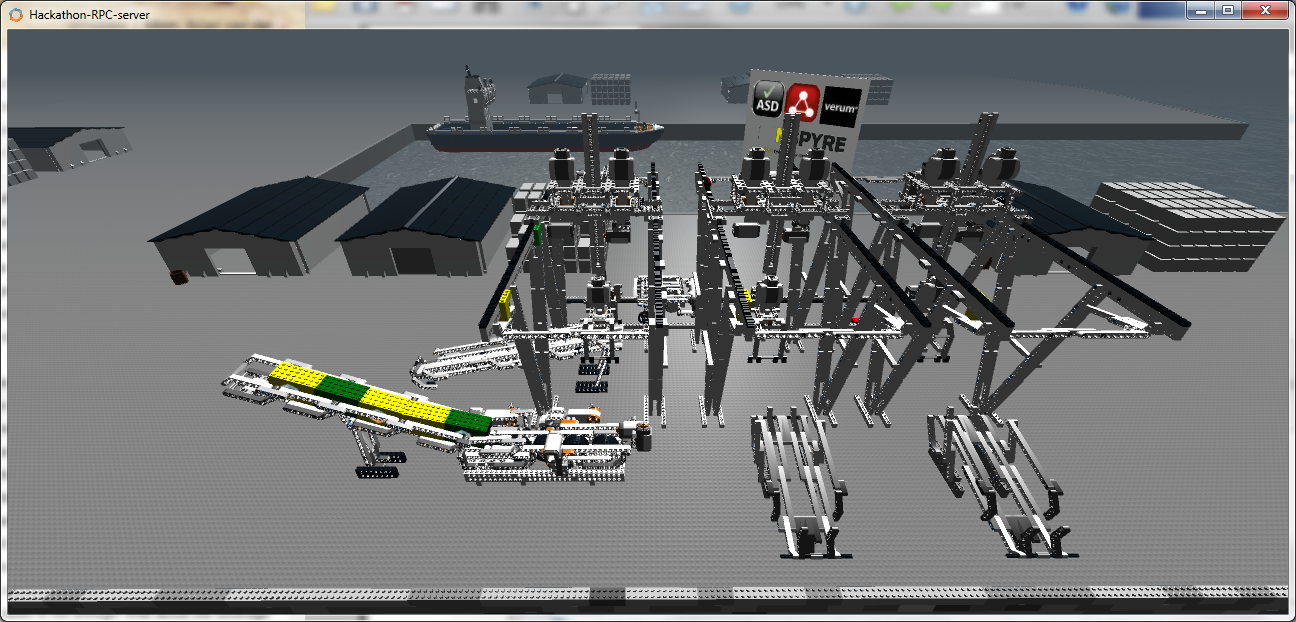}
\caption{Container Terminal Global Structure}
\label{fig:ContainerPicture}
\end{center}
\end{figure}

\begin{figure}[htp]
\begin{center}
\includegraphics[width=0.5\textwidth]{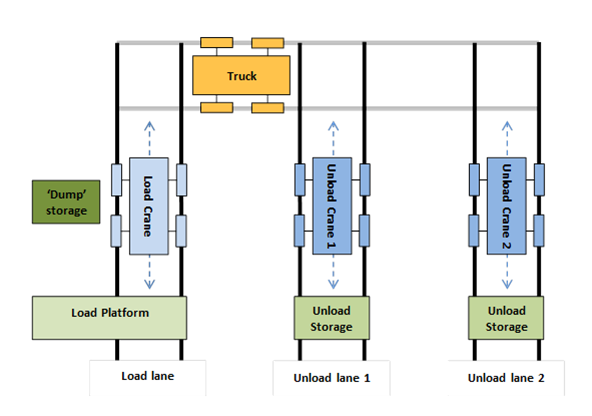}
\caption{Container Terminal Global Structure}
\label{fig:ContainerSchema}
\end{center}
\end{figure}

An overview of the container terminal in question is shown in as a visualization in Figure~\ref{fig:ContainerPicture}, and in schematic form in Figure~\ref{fig:ContainerSchema}. As shown in the pictures, the terminal has three cranes, two of which, situated on the right, are primarily used for loading and unloading containers from the vessel(s), and one, situated on the left, that is used to move the cargo from or to other forms of transport. Goods are transported between cranes by an automated truck that is also part of the system. The main software component in the case is the controller, that has to move container to or from the right cranes in the right order. The system is judged to be working correctly when containers are moved to the right place safely and efficiently. For example, when moving a container to the truck, a crane should not release it until the vehicle is in position to support it, otherwise the container would fall to the ground. 

\subsection{Results}
\label{sec:Results}
Based on the case study, we have drawn several conclusions on the approaches discussed in this paper, which are summarized in Table~\ref{tbl:CaseResults}.
\begin{description}
\item[Approach] General description of the verification approach
\item[Technique] ASD uses model checking to verify design models before code generation. In contrast, MBT creates tests based on manually created test models, that verify actual implementations by executing sequences of commands. \ASDSpec constructs basic test models based on ASD models, which can be used directly to create compliance tests for components or extended to describe requirements not covered by ASD verification, such as those that are data-testing related~\cite{DBLP:conf/issre/VishalKKM12}.
\item[Modeling] ASD uses two kinds of models, interface models and design models. Interface models define behavior that can be provided by a component, and design models
\item[Effort] Because ASD design models need to cover every aspect of the implemented components, a significant effort is required to complete them. In return, ASD can save effort in the project as a whole through early design verification and code generation. In contrast, test models only need to cover the details needed for constructing suitable tests, so they are cheaper to construct, but they are useful only for testing. \ASDSpec automatically generates test models based on existing modes, minimizing effort needed specifically for testing by reusing work that has been done earlier.
\item[Cost] In the original case study, the cost of modeling and constructing the MBT test cases and the test environment was estimated at 28 hours. Because we reused  some results from that work, we cannot directly compare the time spend on the \ASDSpec models with this figure, but as a conservative estimate, we computed a cost of 16 hours to reconstruct the test environment and the test models from scratch. This estimate includes overhead costs based on the original case study, which we expect could actually be reduced using \ASDSpec, further reducing the cost of test creation. 
\item[Tool support] As can be seen in Sections~\ref{sec:ASD} and \ref{sec:SpecExplorer}, ASD models are constructed using a table-based method, and Spec Explorer are defined by annotated \csharp code. Because the ASD language is simpler and more structured, it is easier to define models than in Spec Explorer, at a cost in flexibility. Additionally, the ASD tooling automatically checks model completeness and consistency, which means modeling errors can be detected at an earlier stage than in Spec Explorer. \ASDSpec generates test models, which means the only complexity lies in any test customizations that are added later.
\item[Model composition] In software engineering, decomposition is a common method to reduce the complexity of systems. Some system-wide properties, however, can only be verified by examining all the parts of a system together. Purely component-based approaches, like ASD, are limited in their ability to handle these properties, because only interface models can be explicitly combined in design models, design models cannot be composed for verification purposes. In contrast, MBT test models can be combined to create test cases for entire systems, allowing specific test cases to be created when desired. Because \ASDSpec is based on Spec Explorer, all model composition techniques available in that tool can be used with and added to our generated basic models.
\item[Number of test cases generated] In order to give an indication of the time needed to execute the verification, we look at the number of test cases used by each approach. Because ASD uses static verification only, there are no actual test cases involved. Instead, the time needed for verification depends primarily on the size of the state space, which is related only indirectly to the number of test cases. MBT does use test cases, and for our case study Spec Explorer generated 89 tests, based on combining the default \texttt{shorttests} and \texttt{longtests} test generation strategies and a manually created Spec Explorer model. If we apply the same strategies to the basic test models generated by ASDspec, we get 93 tests. While this number is likely to increase when modifications are made for specific requirements or model composition, it is an indication that the generated models initially posses a similar level of detail as the manually constructed Spec explorer models.
\end{description}
\subsection{Case Study Conclusions} 
Based on our experiences in the case study, we conclude that the main advantage of the ASD approach lies in the combination of code generation and design verification, which allows validated components to be constructed at little cost over a direct, unvalidated implementation. However, the approach cannot be applied to (fully) verify existing components or systems containing them. In contrast, MBT requires at significant effort purely for verification, but can be applied to any system. Finally, \ASDSpec also applies to both generated and external components, at greatly reduced costs, but the generated test models are based purely on the interface models, not on the specific properties of the system, limiting the flexibility of the generated tests. The automated approach does guarantee that the complete interface specification is represented in the model, and can thus be covered during tests.

\begin{table}
\begin{tabular}{p{2cm}|p{4cm}|p{4cm}|p{4cm}}
\textbf{Metrices} & \textbf{ASD} & \textbf{MBT} & \textbf{ASDspec}\\
\hline
Approach & Generate deadlock-free code of components & Generate and execute test suite from behavior model & Generate test model/test suite from ASD interface model\\
\hline
Technique & ASD model checking & Model based testing & Static testing\\
\cline{2-4}
 & Mathematical proof & Verification & Dynamic testing\\
 \cline{2-4}
 & Static testing & Dynamic testing & \\
\hline
Modeling & Create interface model + design model & Create test model & Generate test model form ASD interface model / generate from generated test model\\
\cline{2-4}
 & Based on model decomposition & Supports model composition & Combines decomposition and composition\\
\hline
Effort & High & Medium & Low\\
\cline{2-4}
 & Manually constructed models describe complete system& Manually constructed models describe relevant interfaces  & Generated models describe relevant interfaces\\
\cline{2-4} 
Cost(hrs)  & - & 28 & 16\\
\cline{2-4}
Tool support  & Medium & Low & High\\
\cline{2-4}
 & Specialized table-based modeling language with strong tool support& Combination of two languages, need connection with system to be tested & Generated models, CordScript knowledge needed for test definition\\
\hline
\hline
Model composition & Interface model only & All test models & All test models\\ 
\hline
Number of test cases generated & - & 89 & 93 \\
\end{tabular}
\caption{Key Case Study Results}
\label{tbl:CaseResults}
\end{table}
\section{Related Work}
\label{sec:RelatedWork}
Software verification through both static and dynamic testing is a wide area of research, and we will not cover all of it here. Instead, we focus on the tools used in this paper: ASD::Suite and Spec Explorer. Starting with the first, there have been several papers on ASD and its use in industrial settings, for example~\cite{DBLP:conf/fm/Broadfoot05,DBLP:journals/entcs/HopcroftB05}. More interesting to us is~\cite{6207775}, where ASD is combined with the model checking tool Uppaal to provide more complete verification, like we do here with Spec Explorer. In contrast with our approach, both ASD and Uppaal are based on static verification. The authors argue that Uppaal can handle more generic properties than ASD, in particular in the timing domain. This means the main benefit of combining the two tools lies in more detailed static verification of modeled systems, while our approach attempts to widen the scope of verification. Another contrast with our approach is that the transformation from ASD to Uppaal is done by hand, while our approach is automatic. The authors do mention automation as a possible future extension of their work, because the manual procedure used is not very complex. Overall, our approach has the advantage that we use dynamic as well as static testing, allowing us to verify components that are not completely modeled, but for which implementations and interface models are available.
In the same way, there have been several papers published on Spec Explorer. Most of these, like \cite{DBLP:conf/fm/CampbellGNSTV05} and \cite{DBLP:conf/sigsoft/VeanesCST05} focus on specific kinds of systems or testing strategies, while we focus on test model creation.  To our knowledge, combining Spec Explorer with other tools is a novel approach that has not been described before. We believe that the ability to apply the functionality Spec Explorer provides, for example in the area of data combinations and model composition, make it worthwhile to extend the kind of projects it can be used for by creating connections with other tools.
 
\section{Conclusions}
\label{sec:Conclusions}
In this paper, we have described a novel approach to verification that is based on combining model checking and model-based testing. To compare this approach with pure model checking and pure model based testing, we applied it to an existing case study of a container terminal. For model checking we use the ASD::suite tools both to perform static testing and code generation, doing no verification after the code has been generated. This approach has the lowest development cost of the three, while still offering some guarantees that the system meets all requirement, but it relies heavily on the assumption that external code implements interfaces correctly. The second approach, model-based testing, is implemented using the Microsoft Spec Explorer tool. In model-based testing, test models are used to generate test cases to verify implemented components and systems. While the extra modeling effort required increases the cost of this approach compared to the first one, components that where previously not verified can now be tested both in isolation and the context of an entire composed system. The third and new approach uses MBT based on models generated by \ASDSpec based on ASD models. In the \ASDSpec approach, we attempt to achieve all advantages of MBT without the extra model creation costs, by generating test models based on ASD interface definitions. In the case study, we confirmed that a significant reduction in cost of the testing process can be achieved. While the \ASDSpec approach is less flexible than direct MBT, we feel it still provides significant verification for a variety of systems by extending ASD model checking with test results. As a next step, we intend to investigate combining advanced features of Spec Explorer with our tool, to further extend the verification possible, and to develop connections with other MBT frameworks, to access their unique features.
\bibliographystyle{eptcs}
\bibliography{MBT}
\end{document}